\documentclass[%
 reprint,
superscriptaddress,
showpacs,preprintnumbers,
 amsmath,amssymb,
 aps,
]{revtex4-1}
\usepackage{graphicx}
\usepackage{dcolumn}
\usepackage{bm}
\usepackage{color}
\usepackage[%
    colorlinks=true,
    pdfborder={0 0 0},
    linkcolor=red
]{hyperref}
\renewcommand{\thefigure}{\arabic{figure}}
\def\be{\begin{equation}}
\def\ee{\end{equation}}

\begin{document}
\renewcommand{\thefigure}{\arabic{figure}}

\title{ Topological phases in $\alpha$-Li$_{\rm 3}$N-type crystal structure of light-element compounds}
\author{Ali Ebrahimian}
\email{aliebrahimian@ipm.ir}
\affiliation{School of Nano Science, Institute for Research in Fundamental Sciences (IPM), Tehran 19395-5531, Iran}
\author{Reza Asgari}
\email{asgari@ipm.ir}
\affiliation{School of Nano Science, Institute for Research in Fundamental Sciences (IPM), Tehran 19395-5531, Iran}
\affiliation{School of Physics, Institute for Research in Fundamental Sciences (IPM), Tehran 19395-5531, Iran}
\affiliation{ARC Centre of Excellence in Future Low-Energy Electronics Technologies, UNSW Node, Sydney 2052, Australia}
\date{\today}
\author{Mehrdad Dadsetani}
\affiliation{Department of Physics, Lorestan University, Khoramabad 68151-44316, Iran}

\begin{abstract}
Materials with tunable topological features, simple crystal structure and flexible synthesis, are in extraordinary demand towards technological exploitation of unique properties of topological nodal points. The controlled design of the lattice geometry of light elements is determined by utilizing density functional theory and the effective Hamiltonian model together with the symmetry analysis. This provides an intriguing venue for reasonably achieving various distinct types of novel fermions. We, therefore, show that a nodal line (type-I and II), Dirac fermion, and triple point (TP) fermionic excitation can potentially appear as a direct result of a band inversion in group-I nitrides with $\alpha$-Li$_{\rm 3}$N-type crystal structure. The imposed strain is exclusively significant for these compounds, and it invariably leads to the considerable modification of the nodal line type. Most importantly, a type-II nodal loop can be realized in the system under strain. These unique characteristics make $\alpha$-Li$_{\rm 3} $N-type crystal structure an ideal playground to achieve various types of novel fermions well-suited for technological applications.
\end{abstract}

\maketitle

\section{Introduction}
Condensed matter systems represent a paradigm for studying Dirac (Weyl) physics and realize novel phases which have no counterpart in high-energy physics ~\cite{1, 2}. The low-energy excitations of band crossings in Dirac semimetals are described as a four-component Dirac fermion which is carefully composed of two Weyl fermions of opposite chirality. These zero-dimensional nodal points are fourfold degenerate appropriate to the presence of time-reversal symmetry (TRS) and inversion symmetry (IS).
Two bands may also intersect linearly along a one-dimensional manifold in the reciprocal space~\cite{3, 4, 5, 6, 7} of nodal line semimetals (NLSs). Near to such a band crossing point, the tilt of the linear spectrum can be typically used to correctly classify the nodal points. For type-I nodal points, the two crossing bands are slightly tilted and the electron-like and hole-like states touch at a point-like Fermi surface whereas, for type-II points, the two crossing bands are strongly tilted and tipped over along one transverse direction; leading to distinct magnetic and transport responses~\cite{8, 9, 10}.
Furthermore, recent discoveries promptly confirm the existence of threefold (triply), sixfold, or eightfold degenerate symmetry protected nodal points in topological materials~\cite{1, 2} which can endow these materials with unusual physical properties like large negative magnetoresistance~\cite{11} and unconventional quantum Hall effects ~\cite{12}.

To date, topological states of quantum matter are mostly discovered in materials typically containing heavy elements where the band inversion is scientifically based on the largeness of the spin-orbit coupling (SOC). However, the experimental realization of NLS (topological semimetals) is greatly restricted by spin-orbit interactions. Therefore, it is of considerable interest to design spinless systems, i.e., systems with negligible the SOC, with band inversion fulfilling the condition for the non-triviality of band topology.
The controlled design of the lattice geometry of light elements is suitable for achieving various types of novel fermions; involving an adjustable interplay of orbital interactions, element concentration, and lattice strain. In other words, the band inversion mechanism can be tuned by adjustable parameters like tensile strain (crystal field) to a range of critical values such that band inversion occurs between the valence and conduction bands with opposite parities~\cite{13}. The Dirac semimetal realized in A$_{\rm 3}$Bi (A = Na, K, Rb) properly belongs to this class where the reversal of the band ordering between the conduction and valence bands is not ascribed to the SOC~\cite{14}. The appearance of the Dirac feature in Na$_{\rm 3}$Bi can be carefully controlled by Bi concentration and lattice geometry ~\cite{14, 15}.

In the same way, we find that the hexagonal lattice containing the alkali metal (Li, Na) and Nitrogen (N) atoms are suitable for achieving various types of novel fermions by tuning the lattice strain and composition. In this paper, we show that a nodal line (type-I and II), Dirac fermion and triple point (TP) fermionic excitation (as a topologically protected crossing point of three bands) can potentially appear as the desired result of a band inversion in group-I nitrides with $\alpha$-Li$_{\rm 3}$N-type crystal structure, such as Li$_{\rm 3}$N, Na$_{\rm 3}$N and related alloys Li$_{\rm 2}$NaN. Furthermore, we sufficiently demonstrate that superionic conductor Li$_{\rm 3}$X (X = P, As) can show the same results under lattice strain. Although Lithium nitride (Li$_{\rm 3}$N) as a distinguished superionic conductor with several attractive properties and potentials for possible uses ~\cite{16, 17} has been studied extensively, research on its topological properties is limited.
We show that a crystal lattice extension along the {\bf c} axis via an elastic strain or by carefully replacing Li-ion on top of the N atom with the heavier alkali metal reduces the interactions between N-p$_{\rm z}$ and alkali metal-s orbitals. This leads to the reversal of the band ordering between the conduction and valence bands at the center of the Brillouin zone (BZ) of $\alpha$-Li$_{\rm 3}$-type crystal structure. Having possessed the heavier alkali metal, the band inversion is therefore realized in the Na$_{\rm 3}$N in the absence of the lattice strain.
 
Our investigation based on density functional theory (DFT) calculations and the effective low-energy model Hamiltonian together with the symmetry analysis reveal the possible existence of the triply degenerate nodal point (TDNP); formed by crossing point of three bands where two of which are degenerate along a high-symmetry direction in the momentum space. This TDNP is located near the Femi Level and does not overlap with any other bands in the momentum space providing a unique platform to the definitive experimental study of the triply degenerate nodal point. Interestingly, the electronic band structure of $\alpha$-Li$_{\rm 3}$N-type crystal hosts a type-I nodal loop centered around the $\Gamma$ (A) point and along the K-H high symmetry line in the BZ, protected by the coexistence of the TRS, IS and crystal symmetries. Furthermore, under compressive lattice strain, the type-I nodal loop in Na$_{\rm 3}$N can be transformed into a type-II loop.
Most topological materials transform into typical materials when the material dimensions are varied from bulk structure to a two-dimensional (2D) thin film. More excitingly, the 2D structure of Li$_{\rm 3}$N (Li$_{\rm 2}$N monolayer) hosts the topologically nontrivial electronic states and we find that type-II nodal loop can be realized in monolayer (thin-film) of Li$_{\rm 3}$N. The nodal loop emerges from the crossing between two electron-like bands around the $\Gamma$ point and the highly disperse and flat bands appear in the vicinity of the Fermi level providing the necessary condition for possible high superconducting transition temperature in Li$_{\rm 2}$N monolayer~\cite{18}. It is worth mentioning that a robust pinhole-free-Li$_{\rm 3}$N solid thin-film has been recently fabricated with an excellent performance in Li metal anodes~\cite{19}.
Therefore, it is reasonable to naturally obtain this topological feature in experiments. Our work not only puts forward a novel class of light materials which allows the coexistence of various types of novel fermions but also offers an approach to search for new topological semimetals.

This paper is arranged as follows. In Sec. II, we briefly introduce the simulation methods. Sec. III is devoted to the numerical results of the study, focusing on the electronic and topological properties of the $\alpha$-Li$_{\rm 3}$N-type crystal structure and the Li$_{\rm 2}$N monolayer. Last but not least, we summarize our results in Sec. IV.

\begin{figure}[t]
	\centering
	\includegraphics[width=1.0\linewidth]{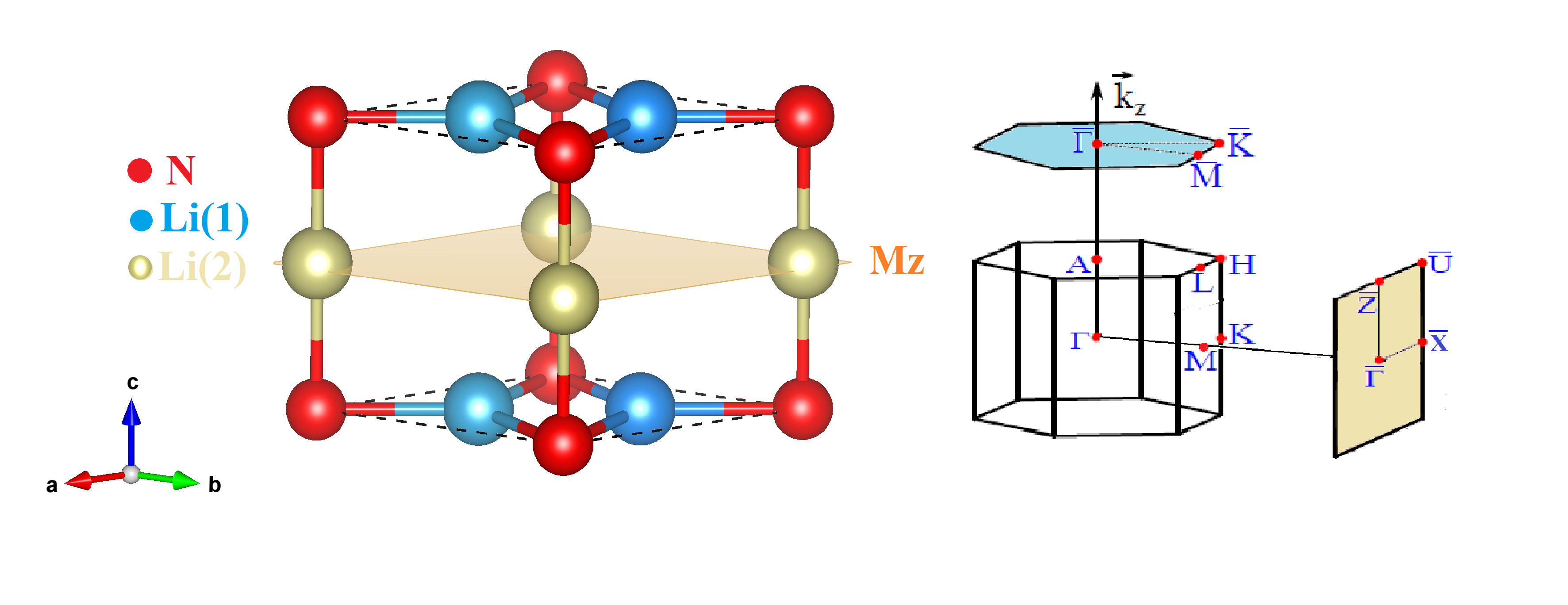}\\
	\caption{(Color online) Left panel: The crystal structure of $\alpha$-Li$_{\rm 3}$N with P6/mmm symmetry. The mirror reflection plane is shown by M$_{\rm z}$. Right panel: The Brillouin zone of the bulk and the projected surface Brillouin zones of the (001) and (100) surfaces.
		\label{fig:1}}
\end{figure}

\section{Theory and simulation methods}

The density functional theory implemented in WIEN2k code~\cite{20} as well as the FHI-aims code package~\cite{21} to meet high accuracy requirements are utilized to carefully explore the topological properties of the system. A 26$\times$26$\times$26 and 30$\times$30$\times$1 Monskhorstpack k-point mesh are used in the bulk and monolayer computations, respectively. The SOC is included consistently within the second variational method and a 20 \AA~ thick vacuum layer is used to prevent interactions between the nearest layers. The WannierTools code~\cite{23} is used to properly investigate the topological properties based on maximal localized functions tight-binding model~\cite{24} that is constructed by using the Wannier90 package~\cite{25} with Nitrogen p and alkali metal s orbitals as projectors. The surface state spectrums are calculated using the iterative Green’s function method~\cite{23, 26}.

\begin{figure*}[t]
\centering
\includegraphics[width=0.88 \textwidth]{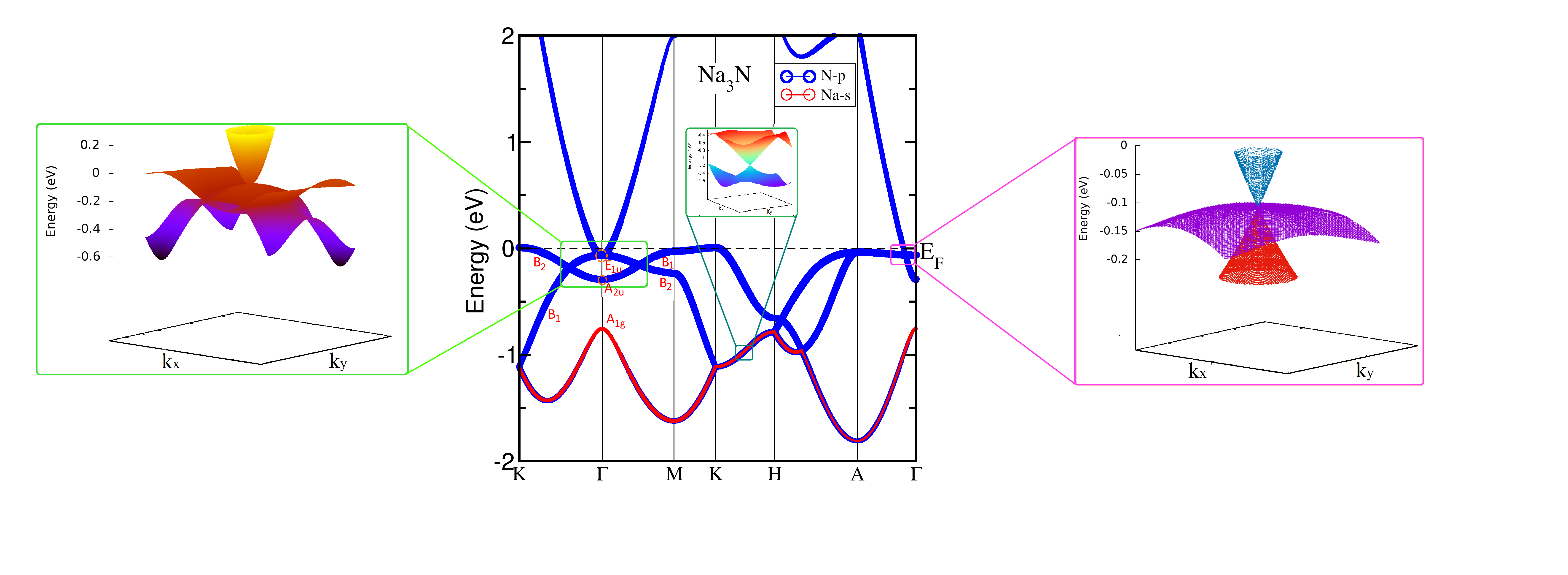} \\
\caption{(Color online) The calculated bulk electronic band structure of Na$_{\rm 3}$N using PBE. The irreducible representation of the selected bands at the $\Gamma$ point and along the $\Gamma$-K and $\Gamma$-M lines are indicated. The orbital-projected band structures of Na$_{\rm 3}$N shows that the band near the Fermi level is mostly composed of Na-s and N-p orbitals. The enlarged 3D plots show the existence of a type-I nodal line around the $\Gamma$ point and along the K-H in addition to TDNP along the $\Gamma$-A.
\label{fig:E}}
\end{figure*}

\section{Results and discussions}
Here as a prominent member of $\alpha$-Li$_{\rm 3}$N structures, we initially consider the electronic and topological properties of Na$_{\rm 3}$N and then consider the Li$_{\rm 3}$N as a well-known insulator of this class of materials to induce the band inversion in this structure through its ternary compounds. Ultimately, we consider the topological properties of monolayer of Li$_{\rm 3}$N.

Lithium nitrite ($\alpha$-Li$_{\rm 3}$N) and Sodium nitrite (Na$_{\rm 3}$N) crystallize in the hexagonal P6/mmm structure ($\alpha$-Li$_{\rm 3}$N-type crystal structure) are shown in Fig.~\ref{fig:1}. The $\alpha$-Li$_{\rm 3}$N structure possesses a layered structure composed of alternating planes of the hexagonal Li$_{\rm 2}$N and pure Li$^{+}$-ions. In the Li$_{\rm 2}$N layers each N (0, 0, 0) is at the center of a regular hexagon formed by the six neighboring Li-ions (1/3, 2/3, 0) and (2/3, 1/3, 0) in units of lattice vectors. From now on, these Li atoms denote as Li(1) and the Li atom above the N atoms will be denoted as Li(2), see Fig.~\ref{fig:1}. In this structure, the three lithium atoms donating their 2s electrons to the nitrogen, result in Li$^{+}$-ions and an N$^{3-}$-ion~\cite{27}. For the Li$_{\rm 3}$N (Na$_{\rm 3}$N) crystal structure, we obtain the lattice constant ${a}$= ${b}$= 3.648 (4.488) and ${c}$= 3.885 (4.660) {\AA} which are completely consistent with those values reported in experiment~\cite{27, 28, 29}.
\begin{figure}[t]
	\centering
	\includegraphics[width=1.0\linewidth]{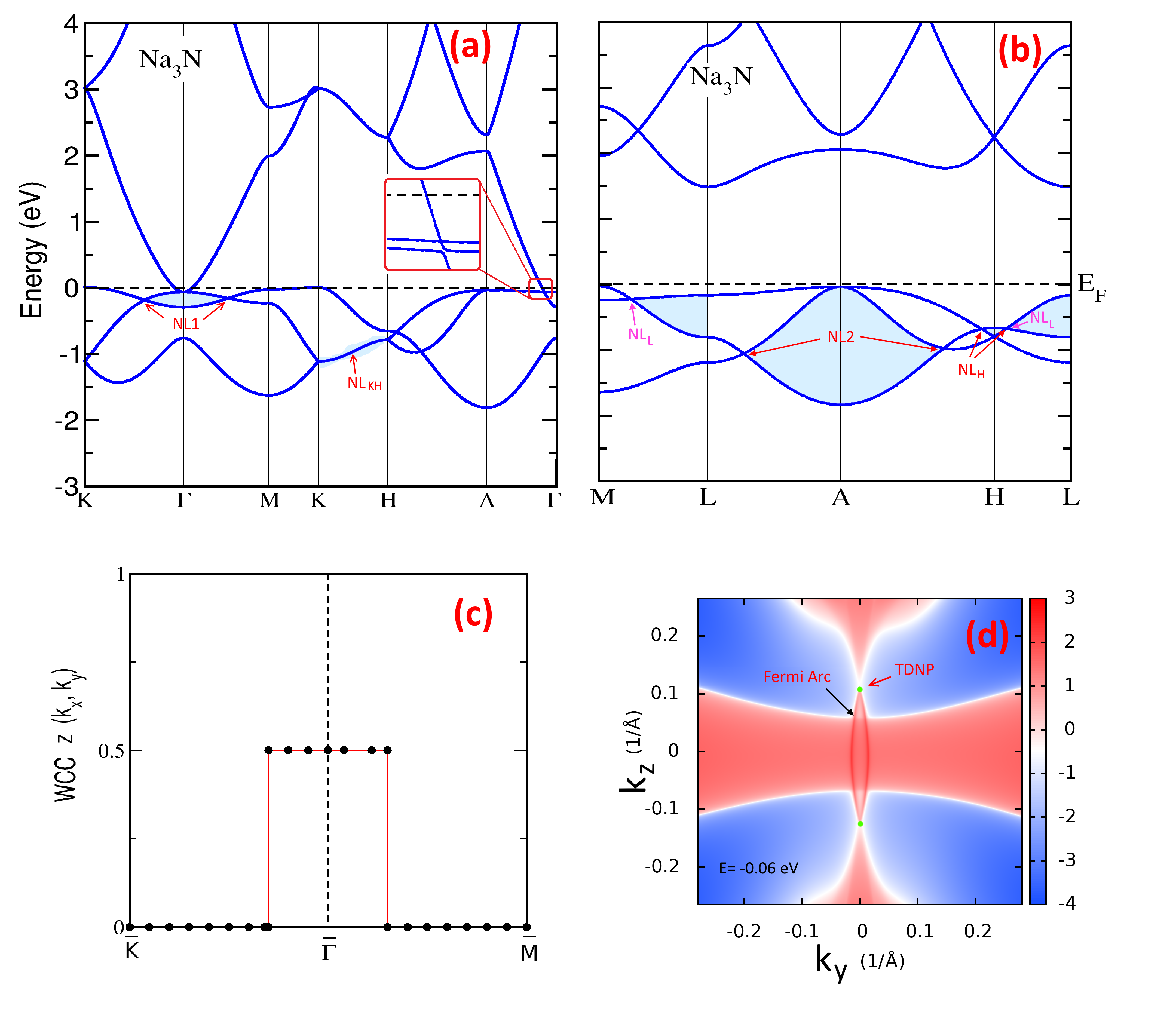}\\
	\caption{(Color online) (a-b) The calculated bulk electronic band structure of Na$_{\rm 3}$N with the SOC. Nl1 (Nl2) shows the nodal line around the $\Gamma$ (A) point at k$_{\rm z}$=0 (k$_{\rm z}$= $\pi$). NL$_{\rm KH}$ shows the nodal line along the K-H, while NL$_{\rm H}$ (NL$_{\rm L}$) shows the nodal line around the H (L) point at (001) ((100)) plane. (c) The evolution of the Wannier charge center z(k$_{\rm x}$ ,k$_{\rm y}$) for the occupied bands of Na$_{\rm 3}$N. (d) The calculated (100) Fermi surface of Na$_{\rm 3}$N with the chemical potential at -60 meV. The red lines (green dots) indicates the Fermi Arc (TDNP).
		\label{fig:4}}
\end{figure}

Figure~\ref{fig:E} shows the orbital resolved band structure of Na$_{\rm 3}$N where the conduction band minimum (CBM) around the $\Gamma$ point is parabolic and mainly originates from N-p orbital while the valence band maximum (VBM) along the $\Gamma$-A represents a doubly-degenerate band and mostly drives from N-p orbital. As shown in the calculated band structure of Na$_{\rm 3}$N (Fig.~\ref{fig:E}), the CBM crosses the doubly-degenerate VBM along the $\Gamma$-A generating a TDNP. Consequently, the Na-s band shifts down the N-P band and the band inversion invariably happens at the $\Gamma$ point. Such a s-p band-inversion leads to topological nontriviality in Na$_{\rm 3}$Bi~\cite{30}.
\begin{figure*}[t]
\centering
\includegraphics[width=0.98\textwidth]{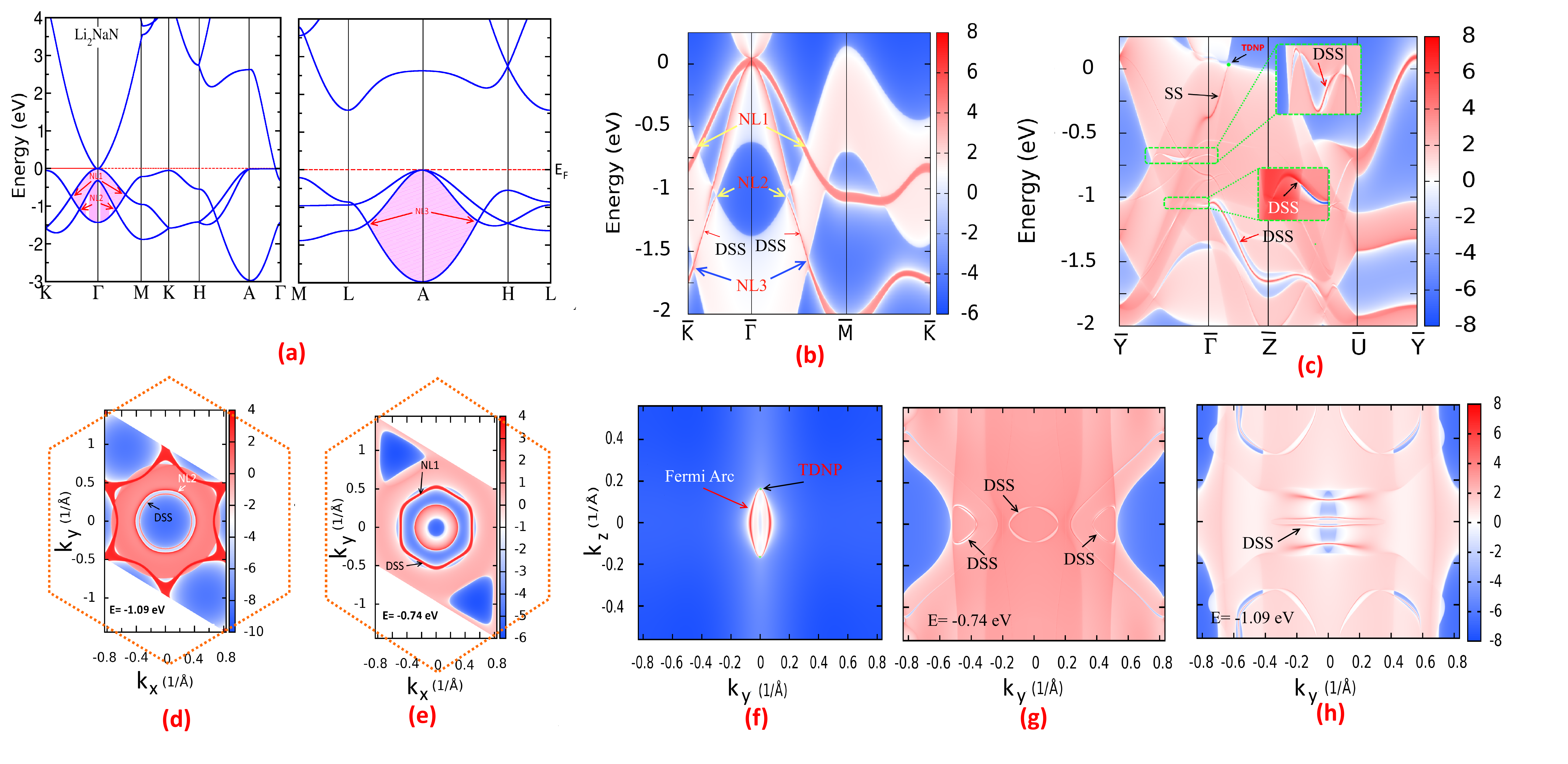}\\
\caption{(Color online) The calculated bulk electronic band structure of Li$_{\rm 2}$NaN. The Nli(i=1, 2) (Nl3) shows the nodal line around the $\Gamma$ (A) point at k$_{\rm z}$=0 (k$_{\rm z}$= $\pi$). NL$_{\rm KH}$ shows the nodal line along the K-H. The calculated (001) (b) and (100) (c) surface band structure of Li$_{\rm 2}$NaN show the existence of the drumhead surface states and nodal lines. The calculated (001) (d-e) and (100) (f-h) Fermi surfaces of Li$_{\rm 2}$NaN with the chemical potential at TDNP and NL1, Nl2. The isoenergy contour shows the nodal-line states and SSs. 
\label{fig:7}}
\end{figure*}

\begin{figure*}[t]
\centering
\includegraphics[width=0.8\textwidth]{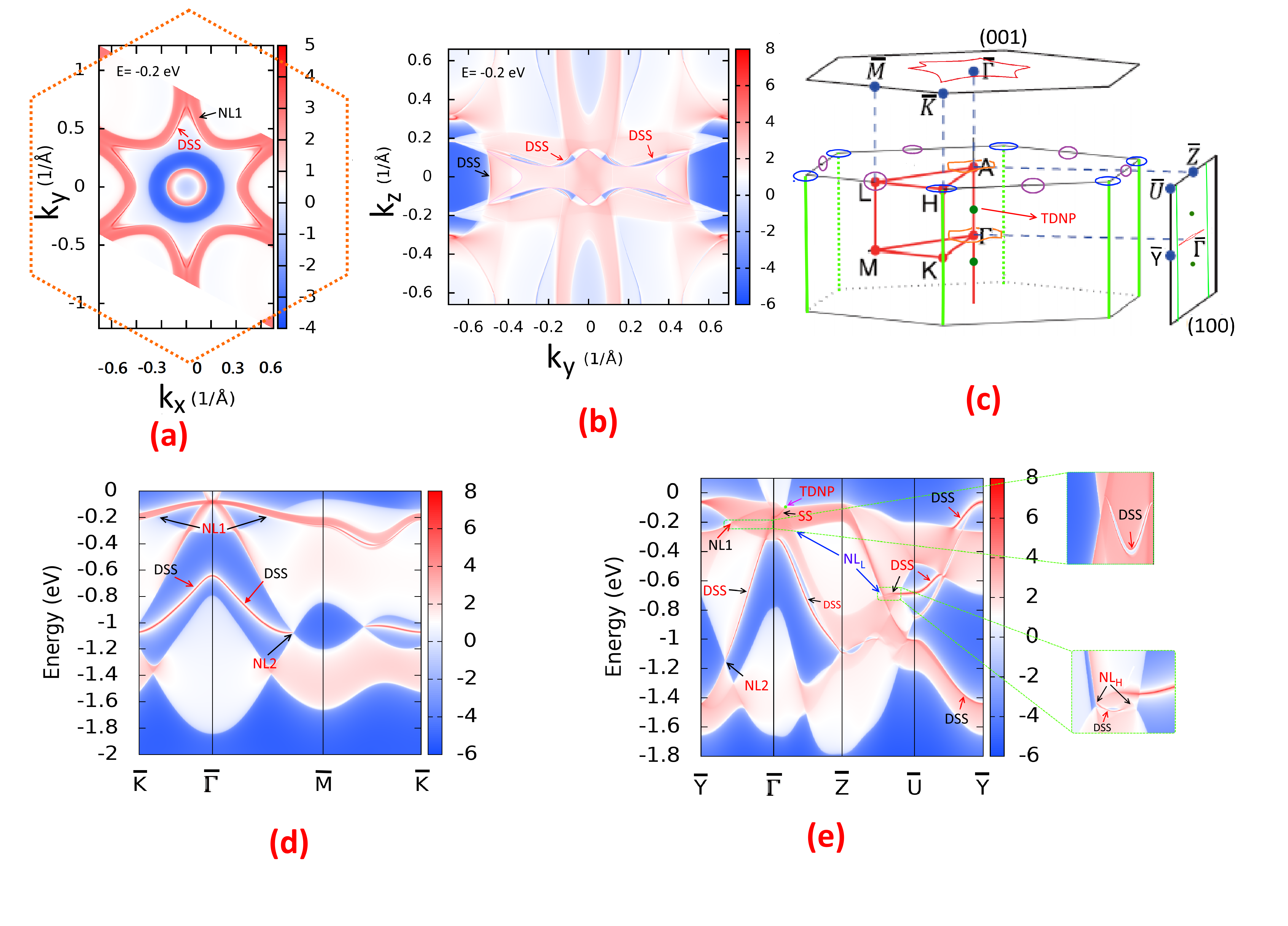}\\
\caption{(Color online) The calculated (001) (a) and (100) (b) Fermi surfaces of Na$_{\rm 3}$N with the chemical potential at -200 meV. The isoenergy contour is showing the nodal-line states (NL) and surface states. (c) The nodal lines and points in the bulk BZ and their projections on the (100) and (001) surface planes in Na$_{\rm 3}$N. The type-I nodal lines NL1 (Nl2) are located inside the bulk BZ at k$_{\rm z}$=0 (k$_{\rm z}$= $\pi$) and marked in brown color. The type-I nodal lines (NL$_{\rm KH}$) are located along the hinges and shown in solid-green color. The type-I nodal lines around the H (NL$_{\rm H}$) and around the L (NL$_{\rm L}$) are marked in blue and purple colors, respectively.  The calculated (001) (d) and (100) (e) surface band structure of Na$_{\rm 3}$N show the existence of the drumhead surface states and nodal lines.
\label{fig:sub}}
\end{figure*}

\begin{figure}[b]
	\centering
	\includegraphics[width=1.0\linewidth]{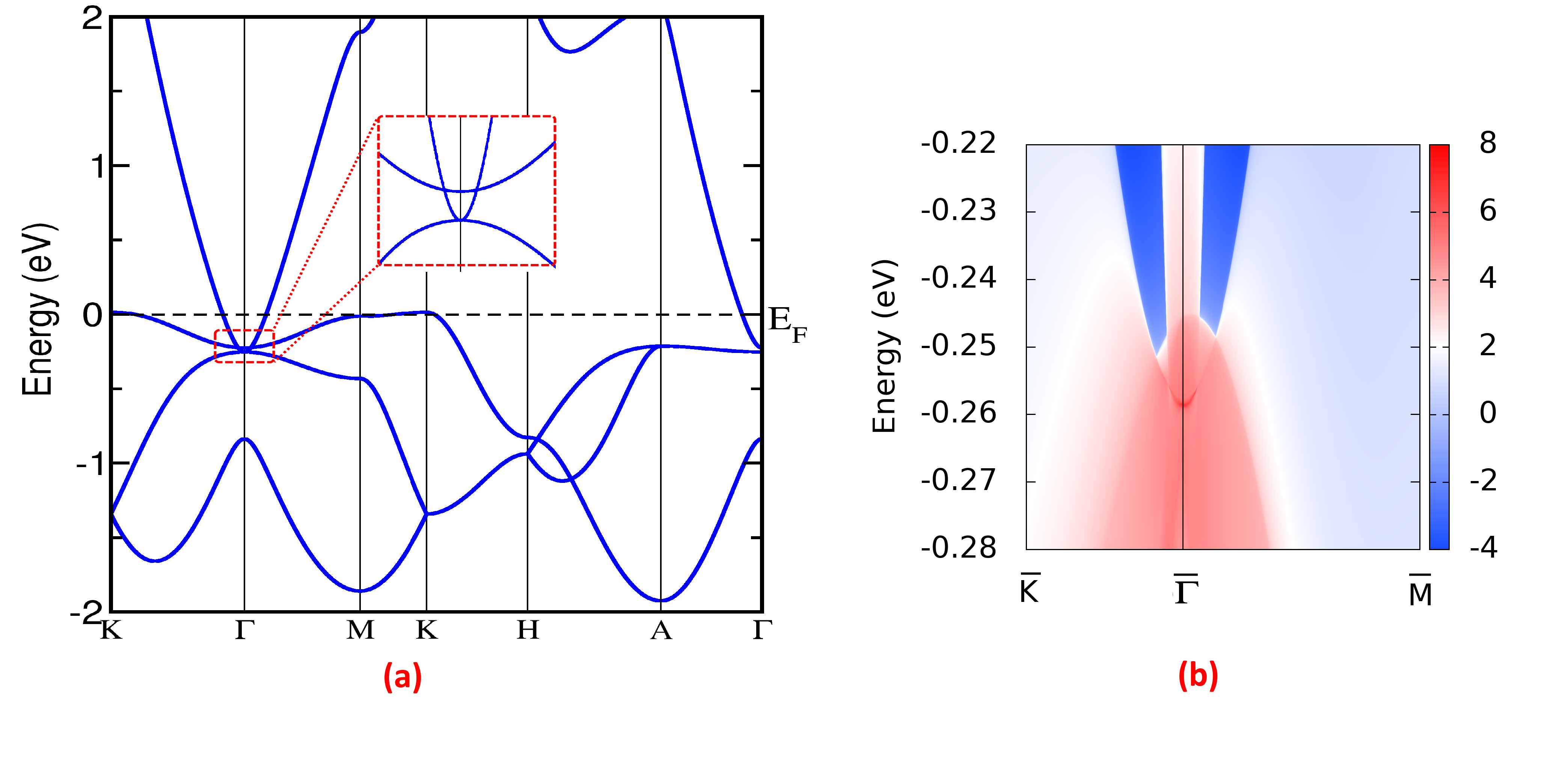}\\
	\caption{(Color online) (a) The calculated bulk band structure of -4\%-strained Na$_{\rm 3}$N. The enlarged plot shows the type-II nodal line around $\Gamma$ point at k$_{\rm z}$= 0. The calculated (001) (b) surface band structure of -4\%-strained Na$_{\rm 3}$N show the existence of the drumhead surface states (DSS) is nestled inside the projected nodal loop.
		\label{fig:8}}
\end{figure}

\begin{figure*}[t]
\centering
\includegraphics[width=0.92\textwidth]{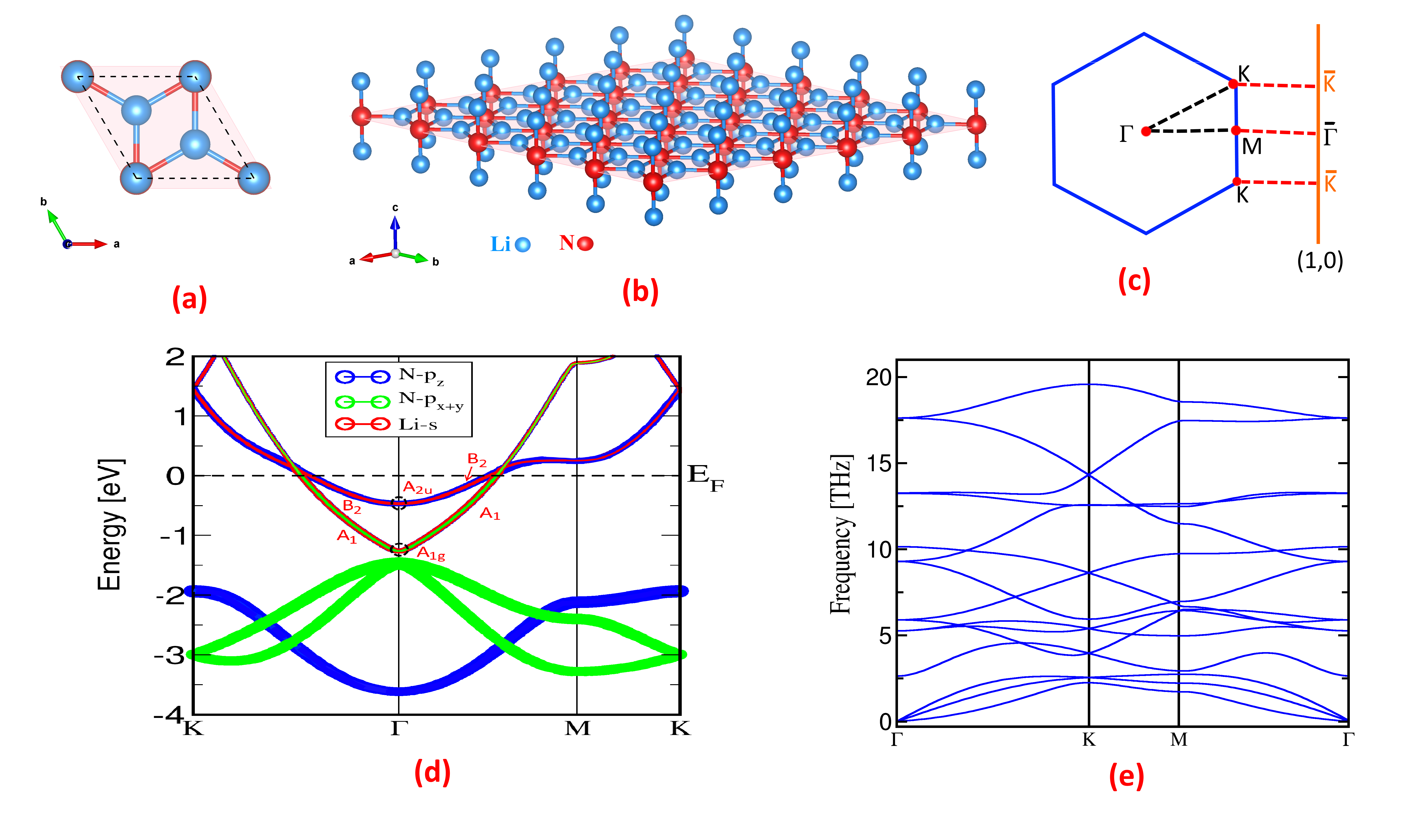}\\
\caption{(Color online) Top view (a) and side view (b) of the crystal structure of Li$_{\rm 2}$N monolayer. (c) The first BZ and (1, 0) edge of Li$_{\rm 2}$N monolayer. (d) The calculated orbital-projected band structure of Li$_{\rm 2}$N monolayer. The irreducible representation of the selected bands at the $\Gamma$ and along the $\Gamma$-K and $\Gamma$-M lines are indicated. (e) The phonon dispersions of optimized Li$_{\rm 2}$N monolayer.
\label{fig:9}}
\end{figure*}

Interestingly, this band crossing in the Na$_{\rm 3}$N forms an ideal TDNP along the $\Gamma$-A very close to the Fermi level without overlapping with any other bands in the momentum space. In addition, this adequately providing a unique platform to the exhaustive experimental study of the TDNP and the physical mechanism for the band inversion in $\alpha$-Li$_{\rm 3}$N-type crystal structures.
Moreover, there are two bands crossing along the $\Gamma$-K and $\Gamma$-M making a nodal ring (NL) around the $\Gamma$ point of Na$_{\rm 3}$N. As shown in Figs.~\ref{fig:E} and~\ref{fig:4} the type-I nodal lines (NL$_{\rm KH}$) exist along the K-H direction at the hinges of the hexagonal BZ which are under protection of C$_{\rm 3z}$ symmetry.
We further explore the evolution of the 1D hybrid Wannier charge centers (WCCs) along a high-symmetry path in the k$_{\rm x}$-k$_{\rm y}$ plane ~\cite{31, 32} to identify nontrivial topology of the electronic structures in $\alpha$-Li$_{\rm 3}$N-type crystal structure. The 1D hybrid Wannier functions are constructed as~\cite{32}
\begin{equation}
|W_{n{l_{z}}}(k_{x}, k_{y})> = \frac{1}{2 \pi}\int d{k_{z}}e^{i{\bf{k}}.(r-l_{z}c\hat{z})}|u_{n,k}>
\end{equation} 
where $l_{\rm z}$ is a layer index and $c$ is the lattice constant along the $\hat{z}$ direction. The WCC along the localized direction $\bar{z}_{\rm n}$ is defined as the expectation value of the position operator $\hat{z}$ for the hybrid Wannier function in the home unit-cell $\bf {R}$ = 0, $\bar{z}_{\rm n}(k_{\rm x}, k_{\rm y})=<W_{\rm n0}|\hat{z}|W_{\rm n0}>$. The sum $z(k_{\rm x}, k_{\rm y})=\sum_{\rm n} \bar{z}_{\rm n}(k_{\rm x}, k_{\rm y})$  over occupied bands gives a direct way to identify the topology of the whole occupied subspace. 
The total hybrid WCC $z(k_{\rm x}, k_{\rm y})$ is equivalent (up to a factor of 2$\pi$) to the total Berry phase of the occupied Bloch functions accumulated along the k$_{\rm z}$ direction. In the presence of the TRS and IS, the Berry phase for an arbitrary loop in the BZ of the spinless system has to be either 0 or $\pi$~\cite{32}. Therefore, the $z(k_{\rm x}, k_{\rm y})$ is either 0 or 1/2 at any $(k_{\rm x}, k_{\rm y})$ and it jumps by 1/2 when passing through a projected nodal point. For Na$_{\rm 3}$N, our calculations show that the $z(k_{\rm x}, k_{\rm y})$=1/2 within the projected nodal loop centered at the $\Gamma$ point in the projected 2D BZ and jumps to zero when passing through the projected nodal loop to the outside (Fig.~\ref{fig:4}). 

We explore that the conduction band along $\Gamma$-A highly depends on the lattice constant value ${c}$ such that the missing band inversion in Li$_{\rm 3}$N is related to its smaller ${c}$ values in comparison to Na$_{\rm 3}$N. More detailed calculations are presented in the Appendix.
 Therefore, in order to possess the reversal of the band ordering between the conduction and valence bands of Li$_{\rm 3}$N, we propose a structure where Li-ion on top of the N atom is replaced by the heavier alkali metal. In fact, a previous comprehensive investigation measured~\cite{33} the band inversion desired range of the lattice constant $ {c}$ in the ternary compounds of Li$_{\rm 3}$N conducting us to obtain topological nontriviality in the Li$_{\rm 2}$NaN (Li$_{\rm 2}$KN).
They indeed showed that the stable Li$_{\rm 2}$NaN crystallizes in hexagonal lattice with the lattice constant ${a}$= ${b}$= 3.65 and ${c}$= 4.60 {\AA}. Therefore, we investigate the possible existence of the nodal points in ternary compound Li$_{\rm 2}$NaN (Li$_{\rm 2}$KN). The lattice constants and internal coordinates of Li$_{\rm 2}$NaN (Li$_{\rm 2}$KN) were fully optimized and the obtained results are completely consistent with the previous report~\cite{33} (Table~\ref{table:Tab}). Our calculations show that replacing Li-ion on top of the N with Na (K) atoms reduces the interactions between the N-p$_{\rm z}$ and Na(K)-s orbitals and leads to the reversal of the band ordering between the conduction and valence bands at the center of BZ of Li$_{\rm 3}$N. As shown in Fig.~\ref{fig:7}, the same as Na$_{\rm 3}$N, the band crossing occurs along the $\Gamma$-A making a TDNP near the Fermi level. Moreover, there are three bands crossing along the $\Gamma$-K and $\Gamma$-M making a nodal ring around the $\Gamma$ point of Li$_{\rm 2}$NaN (Li$_{\rm 2}$KN).

To obtain valuable insight into the protection mechanism for nodal points along the high-symmetry path $\Gamma$-A, $\Gamma$-K and $\Gamma$-M, we analyze the group symmetry and the parity of the bands. The point group of $\alpha$-Li$_{\rm 3}$N is D$_{\rm 6h}$(6/mmm), so the point groups of the wave vector at the high-symmetry point $\Gamma$, K and M are D$_{\rm 6h}$, D$_{\rm 3h}$, and D$_{\rm 2h}$, respectively. This crystal structure naturally possesses three-fold rotational symmetry around the $z-$axis (C$_{\rm 3z}$), the inversion symmetry (IS), the TRS, and the mirror symmetry R$_{\rm z}$= $\sigma _{\rm z}$.
The D$_{\rm 6h}$ point group at the $\Gamma$ point splits the N-p orbitals into two groups (p$_{\rm x}$, p$_{\rm y}$) and p$_{\rm z}$. The point group along the high symmetry line $\Gamma$-K ($\Gamma$-M) is C$_{\rm 2v}$ (mm2) which has four irreducible representations. The system contains vertical mirror plane $\sigma _{\rm v}$, horizontal mirror plane $\sigma _{\rm d}$ (R$_{\rm z}$= $\sigma _{\rm z}$) and C$_{\rm 2}$ rotation axis existing along the K-$\Gamma$ and $\Gamma$-M. The calculated irreducible representations of the point group of the crossing bands along the $\Gamma$-A of Na$_{\rm 3}$N (Li$_{\rm 2}$NaN) show that these bands contain different representations E$_{\rm 1u}$ and A$_{\rm 1g}$ under threefold rotational symmetry, therefore, they cannot interact and mix with each other making a TDNP along the $\Gamma$-A at the Fermi level. The inverted two bands maintain opposite parities with respect to the IS operation.

The two highest valence bands along the K-$\Gamma$ and $\Gamma$-M properly belong to two different 1D irreducible representations preventing them from hybridizing (see Fig.~\ref{fig:E}). Moreover, the two crossing bands around the $\Gamma$ point have opposite eigenvalues with respect to the mirror reflection operation R$_{\rm z}$. Therefore, the intersection of the two bands is protected by the mirror reflection symmetry, providing a continuous closed loop on the mirror plane k$_{\rm z}$ = 0. This mirror reflection protection can be further considered by artificially tuning the structure to remove the mirror reflection symmetry R$_{\rm z}$. This reflection symmetry can be broken by transforming one of the atoms slightly along the $z-$direction ($\bf {c}$ axis). As shown in the Appendix, by breaking the reflection symmetry, two crossing bands around the $\Gamma$ are found to belong to the same representation of the reduced space group, and the band-crossing is gapped. This result confirms that the nodal line ring is protected by the mirror reflection symmetry. In other words, the time-reversal symmetry together with inversion and mirror symmetry guarantee the occurrence of the node line in $\alpha$-Li$_{\rm 3}$N-type crystal structures~\cite{34, 35, 36}.

\begin{figure*}[t]
\centering
\includegraphics[width=0.92\textwidth]{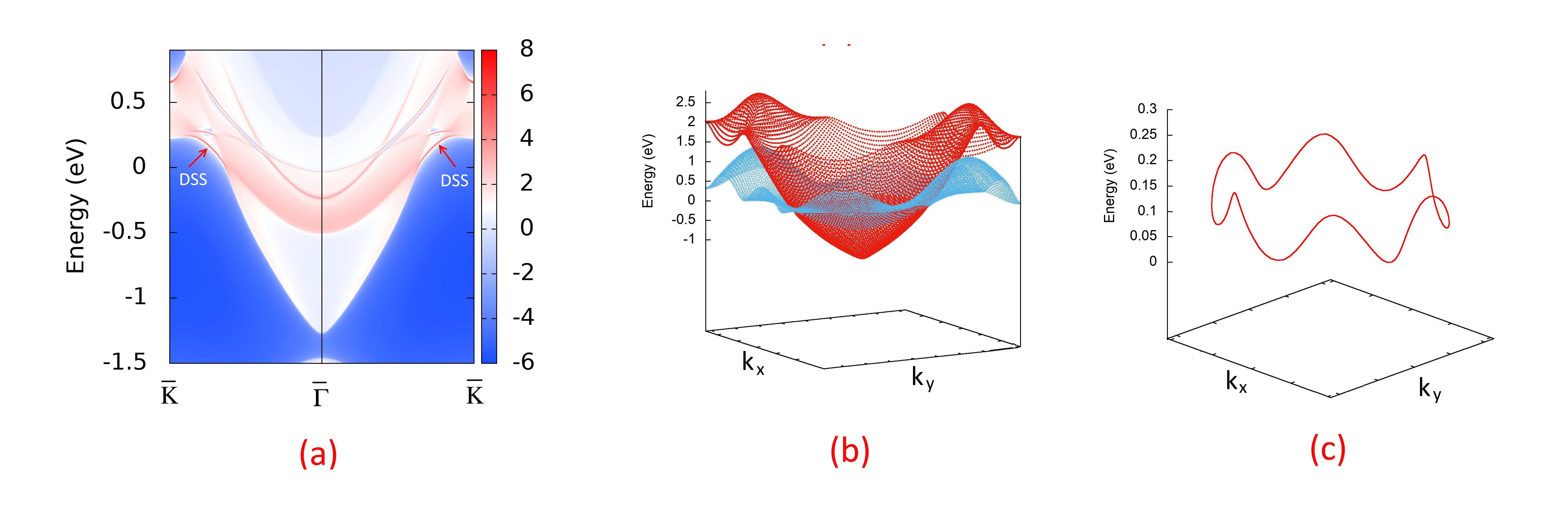}\\
\caption{(Color online) (a) The calculated (1, 0) edge band structure of Li$_{\rm 2}$N shows the existence of the DSS which is nestled outside the projected nodal loop. (b) The enlarged 3D plots show the existence of the type-II nodal line around the $\Gamma$.  (c) The calculated nodal line in the E-k$_{\rm x}$-k$_{\rm y}$ space using the effective Hamiltonian of Eqs. (4) and (6).}
\label{fig:10}
\end{figure*}

From the electronic band structure in Figs.~\ref{fig:4} and ~\ref{fig:7}, we can realize that $\alpha$-Li$_{\rm 3}$N-type crystal structure hosts nodal ring centering the A point in the k$_{\rm z}$= $\pi$ plane. Since the band structure in k$_{\rm z}$= $\pi$ plane is relatively similar to that in k$_{\rm z}$= 0 plane, the protection mechanism of the band crossing is the same as mentioned above. Furthermore, Fig.~\ref{fig:4} shows that the Na$_{\rm 3}$N structure emerges two more type-I nodal rings N$_{\rm LH}$ (centering the H point in the k$_{\rm z}$= $\pi$ plane) and NL$_{\rm L}$ (surrounding the L point in k$_{\rm z}$= $\pi$ plane) which are efficiently generated by accidental band crossings.

Having taken the SOC into account, the band structure of Na$_{\rm 3}$N (Li$_{\rm 2}$NaN) remains practically unchanged in compare with that obtained without SOC since the SOC is relatively weak in the light elements and their compounds. Our numerical calculations show that doubly-degenerate VBM of Na$_{\rm 3}$N along the $\Gamma$-A splits into two flat bands with a negligible gap approximately 8 meV. As shown in Fig.~\ref{fig:4}, once the SOC turns on, the crossings of the branches with various representations of the space group remain gapless as a result of the symmetry protection; forming a 3D massless fourfold degenerate Dirac point along the $\Gamma$-A due to the presence of the TRS and IS. The SOC opens a bandgap approximately 6 meV along the nodal line of Li$_{\rm 2}$NaN which is negligible and smaller than the SOC-induced gap opening in the most reported nodal line semimetals~\cite{35}. Accordingly, the SOC effect can be neglected.

A prominent sign of a topological semimetal endures the existence of the nontrivial surface states (SSs) on the surface of this material. To carefully investigate the surface band structure and its corresponding Fermi surface, we calculate the local density of states for the (001) and (100) surfaces. Because of the similar band structures and the topological features, we investigate the SSs of Na$_{\rm 3}$N and Li$_{\rm 2}$NaN in the following.

Figure~\ref{fig:sub} schematically shows the formation of the nodal points and the corresponding projected points in (100) and (001) SSs of the BZ of Na$_{\rm 3}$N. The projections of the nodal rings (NLi=1, 2) around the $\Gamma$ point on the (001) surface overlap each other so the topological SSs related to these rings are more obvious over the (100) surface. For Na$_{\rm 3}$N and Li$_{\rm 2}$NaN, the projected spectrum for the (100) surface and the corresponding Fermi surface show the presence of the Fermi arc (red line) connecting two projected nodal points (Figs.~\ref{fig:4} and~\ref{fig:7}(f)). As a significant characteristic of the topological nodal-line semimetal, the drumhead like (DSS) surface states emerge either inside or outside the projected nodal loop~\cite{36, 37}.

For Na$_{\rm 3}$N and Li$_{\rm 2}$NaN, the surface band from NL1 disperses inside the bulk band gap and forms a drumhead surface state contour (see Figs.~\ref{fig:sub}(e) and~\ref{fig:7}(c)). The isoenergy contour in the vicinity of the NL1 nodal ring of the Na$_{\rm 3}$N (Fig.~\ref{fig:sub}(a)-(b)) and NL1 and NL2 nodal rings of the Li$_{\rm 2}$NaN (Fig. ~\ref{fig:7}(d)-(h)) promptly confirms the possible existence of the nodal ring and surface band as indicated by the red lines. Note that for the NL1 nodal rings of the Li$_{\rm 2}$NaN, the drumhead surface states overlap with the bulk nodal rings. The surface states along the $\bar{U}$-$\bar{Y}$ direction are related to the projection NL$_{\rm KH}$ in (100) plane which connects $\bar{\Gamma}$-$\bar{Y}$ and $\bar{Z}$-$\bar{U}$ symmetry lines. As shown in Fig.~\ref{fig:sub}(e), for Na$_{\rm 3}$N, the drumhead like SSs are nestled outside (inside) the projected nodal loop NL$_{\rm L}$ (NL$_{\rm H}$)in (100) plane. All these characters can be experimentally measurable by angle-resolved photoemission spectroscopy technique (ARPES).
Interestingly, our calculations show that the Na$_{\rm 3}$N under compressive tensile strain realizes a type-II DNL semimetal. 

Figure~\ref{fig:8} shows the electronic band structures of 4\% compressive strained Na$_{\rm 3}$N obtained from the first-principles calculations. For Na$_{\rm 3}$N, we find the compressive strain modifies the interaction between N-p and Na-s orbitals; converting the energies of the bands near to the Fermi level. The result shows that when the lattice constant ${c}$ is decreased by more than 4 percent, Na$_{\rm 3}$N becomes a type-II nodal line. The resultant surface band structure of the (001) surfaces of - 4\%-strained Na$_{\rm 3}$N is depicted along the projected high-symmetry lines of the hexagonal lattice. As expected, topological SSs appear inside the bulk gap near the $\Gamma$ point. These direct results show that the nodal line properties of Na$_{\rm 3}$N can be tuned under strain.

Most topological materials subtly transform into normal materials when the material dimensions are varied from bulk structure to a 2D thin film. However, topological materials preserving nontriviality independent of their dimensional structure could be better suited for technological applications. The results show the existence of type-II nodal line states in the monolayer of lithium nitride. The Li$_{\rm 2}$N monolayer naturally has a similar structure to $\alpha$-Li$_{\rm 3}$N-type crystal composed of Li$_{\rm 2}$N plane sandwiched between two Li atomic layer. In this structure, Li$_{\rm 2}$N plane is a mirror plane of the structure under mirror operation $Rz$ which is perpendicular to the threefold rotation axis (C$_{\rm 3z}$) (Fig.~\ref{fig:9}). The optimized lattice constant of Li$_{\rm 2}$N monolayer is found to be ${a}$ = ${b}$ = 3.68 \AA. Although Li$_{\rm 3}$N solid thin-film has been fabricated on Li metal surface~\cite{19}, we have calculated a set of the phonon dispersions to check the structural stability of Li$_{\rm 2}$N monolayer.

The calculated phonon spectra (Fig.~\ref{fig:9}) indicates no imaginary modes, implying the structural stability of the Li$_{\rm 3}$N monolayer. Note that the phonon dispersions are calculated by using the 2D Coulomb cutoff within DFPT to eliminate the spurious long-range interactions with the periodic copies and correctly account for the long-wavelength of polar-optical phonons in 2D framework ~\cite{38, 39}.
Figure~\ref{fig:9} shows the band structure of Li$_{\rm 2}$N monolayer. Two electron-like conduction bands cross each other along the $\Gamma$-K and $\Gamma$-M directions making a type-II continuous nodal line surrounding the $\Gamma$ point. These band crossings linearly disperse over a substantial energy range along the $\Gamma$-M ($\Gamma$-K) and remain isolated up to about $E = 2$ eV around the Fermi level. The calculation results indicate that low-energy electronic states mainly come from N-p and Li-s states. The three-dimensional (3D) plot for the nodal line (Fig. ~\ref{fig:10}) presents a notably energy-dependent loop around the $\Gamma$ very close point to the Fermi level.

The calculated 3D band structure around the $\Gamma$ point shows that the energy of the nodes varies within the range of 70 to 170 meV along the loop; making it quite flat in $E-($k$_{\rm x}$-k$_{\rm y})$ space (Fig.~\ref{fig:10}). As shown in Fig. ~\ref{fig:9}, the nodal loop does not overlap with other bands at the Fermi level which naturally makes it an ideal platform to study exotic properties of the type-II nodal loop. Most excitingly, we find that this nodal line is still preserved in the presence of the SOC due to a weak SOC effect in Li$_{\rm 2}$N monolayer. Consequently, the Li$_{\rm 2}$N monolayer is nicely suitable for practical applications as a type-II NL candidate. Note that finding an ideal nodal line states immune from the SOC is still one of the main goals of the field~\cite{40, 41}.
To elucidate the protection mechanism of the band crossing points around the $\Gamma$ point, we identify the irreducible representations (IRs) of the energy bands around the $\Gamma$ points (Fig.~\ref{fig:9}). The point group of the $\Gamma$ point in reciprocal space is D$_{\rm 6h}$ and the two crossing bands around the $\Gamma$ point belong to the A$_{\rm 1g}$ and A$_{\rm 2u}$ IRs. Therefore, the two inverted bands maintain the opposite parity under the IS. The D$_{\rm 6h}$ little group at the $\Gamma$ point ensures the presence of the R$_{\rm z}$ mirror plane perpendicular to the C$_{\rm 3z}$ principle axis in addition to the TRS and IS. In a structure respecting the TRS and IS, such an energy inverted bands with opposite parity guarantees the existence of a closed nodal line~\cite{34, 42}.

The two crossing bands along with the K- $\Gamma$ and $\Gamma$- M belong to two different 1D irreducible representations of C$_{\rm 2v}$ point group preventing them from hybridizing. Furthermore, they naturally maintain opposite parities with respect to the R$_{\rm z}$ operation. Therefore, the crossing bands along the loop are protected by the coexistence of the TRS, IS and R$_{\rm z}$ mirror symmetries. Such degeneracy along the high-symmetry path is similar to the case of Na$_{\rm 3}$N bulk structure. Since the system preserves both the IS and TRS symmetries, a quantized Zak phase is expected for a path along the reciprocal lattice vector perpendicular to the BZ edge~\cite{43, 18}. The Zak phase jumps by $\pi$ at the nodal lines scientifically verifying the nontrivial character of the nodal lines. Put differently, we find that the Zak phase along the $\Gamma$-M path is $\pi$ resulting in the appearance of the SSs at the edge of BZ.
Although the Zak phase is well-defined for nodal loops, we further calculate the $\mathbb{Z}$$_{2}$ invariants based on the parity eigenvalues at the parity-invariant points in the reciprocal space~\cite{44}. The parity products for occupied states at two time-reversal invariant momenta (TRIM) of the $\Gamma$, M are calculated to be $-$ and $+$, respectively, which gives $\mathbb{Z}$$_{2}$= 1, accurately indicating the nontrivial topological phase.

To further characterize the nodal loops, we construct a minimal effective Hamiltonian around the $\Gamma$ point. The spinless effective Hamiltonian for the two bands around the $\Gamma$ point takes the general form:

\be\label{dynamical_term}
H(k)= d_{0}(\vec k)\sigma_{0} + d_{x}(\vec k)\sigma_{x} + d_{y}(\vec k)\sigma_{y} + d_{z}(\vec k)\sigma_{z}
\ee
where the 2$\times$2 identity matrix $\sigma_{0}$ and Pauli matrices $\sigma_{i}$(i = x, y, z) operate in the pseudospin space of the two bands that cross near the $\Gamma$ point and k is relative to this point. Without loss of generality, the term $d_{0}$$\sigma_{0}$ is neglected because it produces just an overall energy shift. The point group at the $\Gamma$ contains a threefold rotation C$_{\rm 3z}$ about k$_{\rm z}$, a twofold rotation C$_{\rm 2}$ about the $\Gamma$- K ($\Gamma$- M) axis, the IS and TRS symmetries. As the two crossing bands around the $\Gamma$ are mostly contributed by the Li-s and N-p$_{\rm z}$ orbitals, we can thus write

\be\label{dynamical_term}
C_{3z}=\sigma_{0}, \,\,\, C_{2}=\sigma_{z},\,\,\, P=\sigma_{z}, \,\,\,T=K\sigma_{0} 
\ee
where $|s>$ and $|p_{\rm z}>$ are chosen as basis vectors. These symmetries impose constraints on d$_{\rm i}$(k) (i= x, y, z) and allows the following expressions for d$_{\rm i}$(k)~\cite{45} on the plane k$_{\rm z}$= 0:
\be\label{dynamical_term}
  d_{x}=0,
  d_{y}=a(k_{x}^{2}k_{y}-k_{x}k_{y}^{2}),
  d_{z}= b + c(k_{x}^{2}+ k_{y}^{2}-k_{x}k_{y})
\ee

Furthermore, the two-band Hamiltonian for a system respecting both the TR and IS symmetries can be chosen as real value, leading to $d_{\rm y}=0$. The inverted band around $\Gamma$ leads to bc $<$ 0 which is required condition for the band crossing on the plane k$_{\rm z}$ = 0. The necessary conditions for the band crossing (d$_{\rm z}$(k)=0) shows that the band touching happens along a snakelike ring around the $\Gamma$ point (Fig.~\ref{fig:10}) confirming the existence of the nodal-line structure. We present the calculated momentum-resolved surface density of states along the high symmetry line of in the (1, 0) edge BZ in Fig.~\ref{fig:10}. The drumhead surface state nestled outside the nodal lines is clearly visible, consistent with the calculated nontrivial character of the nodal lines.

\section{Summary and conclusions}
In summary, we have demonstrated the existence of the various types of nodal points in $\alpha$-Li$_{\rm 3}$N-type crystal structure. Our study based on density functional theory calculations and symmetry analysis sufficiently reveal that a crystal lattice extension along the $\bf {c}$ axis causes the reversal of the band ordering between the conduction and valence bands at the center of the Brillouin zone. This leads to a topological nontriviality in $\alpha$-Li$_{\rm 3}$N structure. The crystal lattice extension can be occurred by replacing Li-ion on top of the N atoms with a heavier alkali metal. As a result, a triply degenerate nodal point near the Fermi level, which does not overlap with any other bands in the momentum space, provides a unique platform to the exhaustive experimental study of the triply degenerate nodal point. Moreover, the electronic band structure of $\alpha$-Li$_{\rm 3}$N-type hosts a type-I nodal loop centered around the $\Gamma$ (A) point and along the K-H high symmetry direction in the Brillouin zone. We have properly found that under compressive lattice strain, the type-I nodal loop in Na$_{\rm 3}$N can be transformed into a type-II loop. More excitingly, a 2D structure of Li$_{\rm 3}$N (Li$_{\rm 2}$N monolayer) hosts the topologically nontrivial electronic states and our calculation show that type-II nodal loop can be realized in monolayer (thin-film) of Li$_{\rm 3}$N.

\begin{figure}
\includegraphics[width=0.95\linewidth]{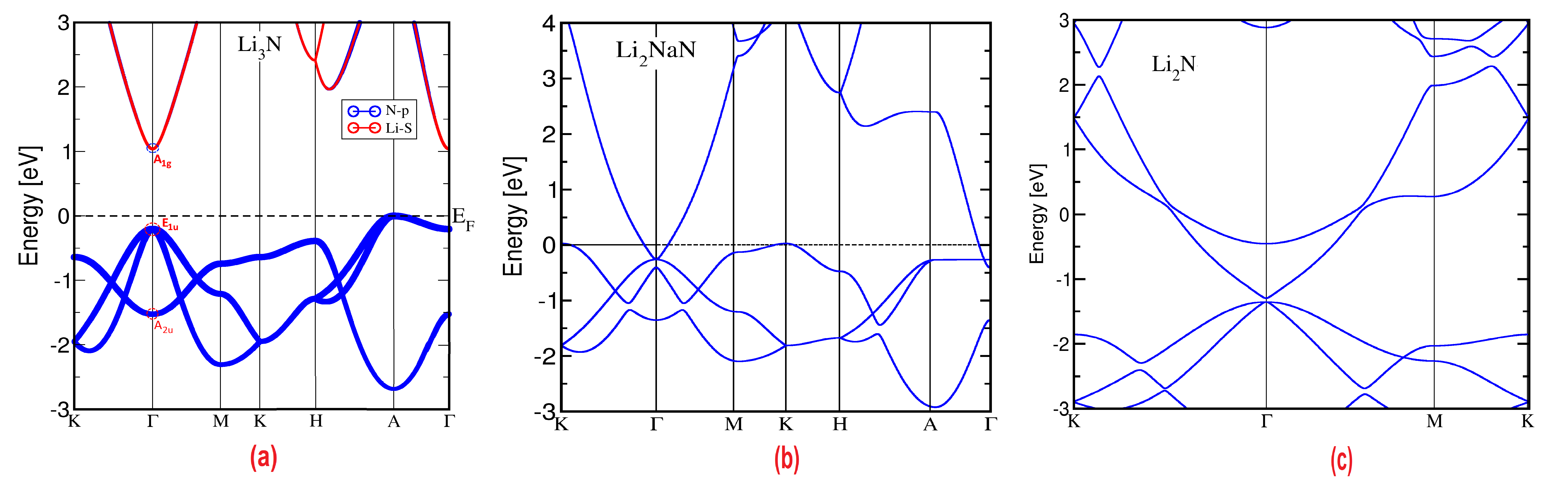}\
\caption{(Color online) (a) Calculated bulk band structure of Li$_{\rm 3}$N using PBE. (b) and (c) The electronic band structure of Li$_{\rm 2}$NaN and Li$_{\rm 2}$N monolayer when the reflection symmetry is broken by moving the Li atom slightly along the $z-$direction.
\label{fig:form}}
\end{figure}

\section{Acknowledgments}

This work is supported by Iran Science Elites Federation. A. E. would
like to acknowledge F. Foolady for her support.

\appendix

\begin{figure*}[!htp]
	\includegraphics[width=0.9\linewidth]{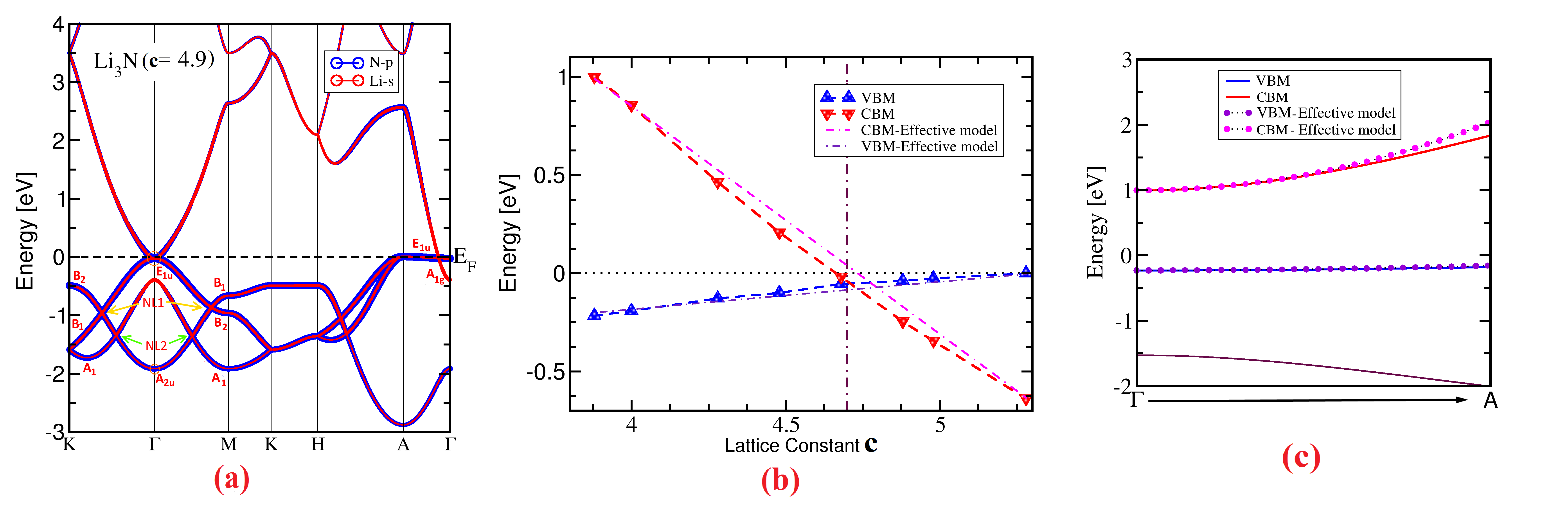}\
	\caption{(Color online) The calculated bulk electronic band structure and the orbital-projected band structures of tensile-strained Li$_{\rm 3}$N. The irreducible representation of the selected bands at the $\Gamma$ point and along the $\Gamma$-A, $\Gamma$-K, $\Gamma$-M lines are indicated. (b) Calculated CBM and VBM of Li$_{\rm 3}$N as a function of the lattice constant $\bf {c}$ using ab initio band structure (dashed line with a triangle) and effective model (dashed line without triangle) (c) The dashed lines show that the effective model reproduces quite well the ab initio band structure (solid blue line) along the $\Gamma$-A.
		\label{fig:hf}}
\end{figure*}
\section{The band structures with more details}\label{B}
Here, we present more details of the band structure calculations of $\alpha$-Li$_{\rm 3}$N-type structure. Figure~\ref {fig:form}(a) shows Calculated bulk band structure of Li$_{\rm 3}$N using PBE. 
Figures~\ref {fig:form}(b) and (c) show the band structure of Li$_{\rm 2}$NaN and Li$_{\rm 2}$N monolayer when the reflection symmetry is broken by moving the Li atom slightly along the $z-$ direction. In spite of the crystal structure similarity between Na$_{\rm 3}$N and Li$_{\rm 3}$N, the CBM and VBM of Li$_{\rm 3}$N are completely separated along the $\Gamma$-A revealing the triviality of this insulator (Fig.~\ref{fig:form}). However, the band structure resemblances of Na$_{\rm 3}$N and Li$_{\rm 3}$N stimulated us to investigate a mechanism for inducing the band inversion in the Li$_{\rm 3}$N structure.
Properly focusing on the possibility of the s-p band inversion, we strain the crystal of Li$_{\rm 3}$N along the {\bf c} axis, while analyzing the band order associated with atomic orbitals. Our calculations show that by increasing the lattice constant {c}, the vertical distance between Li$_{\rm 2}$N layer and Li(2) atom is increased. Therefore, the interaction between N-Li(2) atoms and hybridization between N-p$_{\rm z}$ and Li(2)-s orbitals will be progressively weakened; which lowers (raises) the energy of the CBM (VBM).  The CBM shifts down the VBM above ${c}$= 4.7 \AA ~which can be viewed as a transition to a topological semimetal protected by rotation symmetry, TRS and IS (Fig.~\ref{fig:hf}).
We properly present the low-energy effective Hamiltonian around the band crossing along the $\Gamma$-A; carefully verifying the overall band dispersion of the CBM and VBM obtained from our DFT calculations. Our results show that the ab initio CBM (VBM) along the $\Gamma$-A is reproduced very accurately by this low-energy effective Hamiltonian. Moreover, CBM (VBM) is a function of the lattice constant ${c}$ and can be tuned to induce the band inversion at ${c}$= 4.69. This band inversion mechanism is the same as what happens in Na$_{\rm 3}$N and Na$_{\rm 3}$Bi~\cite{13, 14}. As the band crossing is happened in k$_{\rm z}$= 0 plane with TRS, the 2D system (k$_{\rm z}$ = 0) can be used to define a 2D $\mathbb{Z}$$_{2}$ invariant $\nu$$_{2D}$.
Therefore, the band inversion between the valence and conduction bands with opposite parities (P =$\pm\ \tau_{z}$), changes $\nu$$_{2D}$ by 1 causing nontrivial topological properties in Li$_{\rm 3}$N. Moreover, the structural and topological properties of $\alpha$-Li$_{\rm 3}$N-type structure is given by Table I.
\begin{table}[h!]
        \caption{The structural and topological properties of $\alpha$-Li$_{\rm 3}$N-type structure}
        \begin{tabular}{c  c c  c c} 
                \hline
                \hline
                Crystal    &  Li$_{\rm 3}$N & Na$_{\rm 3}$N & Li$_{\rm 2}$KN & Li$_{\rm 2}$NaN  \\
                \hline
                
               {\bf a}\ ({\AA})  &  3.648 & 4.488 & 3.713 & 3.656    \\
                
                {\bf c}\ ({\AA})& 3.885 & 4.660 & 5.592 & 4.752 \\
                \hline
                 PBE & insulator &  semimetal & semimetal & semimetal  \\
                 $\mathbb{Z}$$_{2}$  & 0 & 1 & 1 & 1 \\              
                \hline
                \hline
        \end{tabular}
        \label{table:Tab}
\end{table}

\section{The low-energy effective Hamiltonian}\label{D}

We present the low-energy effective Hamiltonian around the band crossing along the $\Gamma$-A in strained Li$_{\rm 3}$N; carefully verifying the overall band dispersion of the CBM and VBM obtained from our DFT calculations. In the presence of the SOC the doubly-degenerate VBM of Li$_{\rm 3}$N along the $\Gamma$-A splits into two flat bands with a negligible gap approximately 8 meV. Therefore, in the presence of the SOC, the two crossing bands along the $\Gamma$-A are doubly degenerate due to the simultaneous presence of the TRS and IS. This band crossing can be described by a 4$\times$4 matrix Hamiltonian as a minimal Hamiltonian, and given by~\cite{13}
\be\label{dynamical_term}
{\cal H}(k) = \sum^3_{\rm i,j=0} a_{\rm ij}(k)\sigma_{i}\tau_{j} \begin{pmatrix}
h_{\uparrow \uparrow }(k) & h_{\uparrow \downarrow}(k) \\ h_{\downarrow \uparrow}(k) & h_{\downarrow \downarrow}(k)
\end{pmatrix}
\ee
where the Pauli matrix $\sigma_{1, 2, 3}(\tau_{1, 2, 3})$ indicates the spin (orbital) degrees of freedom and $\sigma_{0}(\tau_{0})$ are the identity matrices. The TRS, IS and 3-fold rotation symmetry about k$_{z}$ (C$_{\rm 3z}$) restrict the possible structure of the minimal Hamiltonian. Along the k$_{z}$ axis, [C$_{\rm 3z}$, ${\cal H}$(k)]=0, so ${\cal H}$(k$_{z}$){$\mid$}$_{\rm k_{x}=k_{y}=0}$ and C$_{\rm 3z}$ can be diagonalized in the same basis. In such a basis, the possible form of IS operator P is restricted to be P =$\pm\ \tau_{0}$, $\pm\ \tau_{x}$, $\pm\ \tau_{z}$ and the Hamiltonian can be written as

\be\label{dynamical_term}
{\cal H}(k_{z}){\mid}_{\rm k_{x}=k_{y}=0}= d_{0}+d(k_{z}, m)\Gamma
\ee
where the $\Gamma$ is either $\Gamma= \tau_{\rm 3}$ or $\Gamma=\sigma_{3} \tau_{\rm 3}$ and and $d(k_{z}, m)$ contain two variables including the momentum component k$_{\rm z}$ and external parameter $m$. In Li$_{\rm 3}$N, along the $\Gamma$-A, the two inverted bands maintain opposite parity and then the inversion operator can be chosen as P =$\pm\ \tau_{z}$.
Therefore, d(k$_{\rm z}$) is even under sign reversal of k$_{\rm z}$ ~\cite{13} and the calculations show that the CBM (VBM) along the $\Gamma$-A is E$_{\rm c}$=~m$_{\rm c}$~-a{k$_{\rm z}$}$^2$ (E$_{\rm v}$=~m$_{\rm v}$~-bk$_{\rm z}$$^2$) in leading order, where a = 12.26 (b= 0.85) and m$_{\rm c}$=~0.99 -1.14 ($\bf{c}$ - 3.88) ( m$_{\rm v}$= -0.2 + 0.14 ($\bf{c}$ - 3.88)). As shown in Fig.~\ref{fig:hf}, E$_{\rm c}$ (E$_{\rm v}$) reproduces very accurately the ab initio CBM (VBM) along the $\Gamma$-A.
Moreover, E$_{\rm c}$ (E$_{\rm v}$) is a function of the lattice constant $\bf{c}$ and can be properly tuned to induce the band inversion at $\bf{c}$= 4.69 (Fig.~\ref{fig:hf}). This band inversion mechanism is the same as what happens in Na$_{\rm 3}$Bi~\cite{13, 14}.
As the band crossing is happened in k$_{\rm z}$= 0 plane with TRS, the 2D system (k$_{\rm z}$ = 0) can be used to define a 2D $\mathbb{Z}$$_{2}$ invariant $\nu$$_{2D}$ . Therefore, the band inversion between the valence and conduction bands with opposite parities (P =$\pm\ \tau_{z}$), changes $\nu$$_{2D}$ by 1 causing nontrivial topological properties in Li$_{\rm 3}$N.


\end{document}